\documentclass[aps,amsmath,amssymb,reprint,superscriptaddress]{revtex4-1}

\usepackage{dcolumn}% Align table columns on decimal point
\usepackage{bm}% bold math
\usepackage{amsmath}
\usepackage{amssymb}
\usepackage{amsfonts}
\usepackage{subfigure}
\usepackage[pdftex]{graphicx}
\usepackage{color}
\usepackage{hyperref}

\usepackage[top=0.75in, bottom=0.75in, left=0.75in, right=0.75in]{geometry}
\begin{document}

%%% Start of article front matter


\title{Measuring behavior across scales}
%Abstracting Behavior: Measuring the Essential Features of The Things Animals Do}

\author{Gordon J. Berman}
\affiliation{Department of Biology, Emory University}
\email{gordon.berman@emory.edu}

%%%%%%%%%%%%%%%%%%%%%%%%%%%%%%%%%%%%%%%%%%%%%%
%%                                          %%
%% Enter the authors here                   %%
%%                                          %%
%% Specify information, if available,       %%
%% in the form:                             %%
%%   <key>={<id1>,<id2>}                    %%
%%   <key>=                                 %%
%% Comment or delete the keys which are     %%
%% not used. Repeat \author command as much %%
%% as required.                             %%
%%                                          %%
%%%%%%%%%%%%%%%%%%%%%%%%%%%%%%%%%%%%%%%%%%%%%%

%%%%%%%%%%%%%%%%%%%%%%%%%%%%%%%%%%%%%%%%%%%%%%
%%                                          %%
%% Enter the authors' addresses here        %%
%%                                          %%
%% Repeat \address commands as much as      %%
%% required.                                %%
%%                                          %%
%%%%%%%%%%%%%%%%%%%%%%%%%%%%%%%%%%%%%%%%%%%%%%

%%%%%%%%%%%%%%%%%%%%%%%%%%%%%%%%%%%%%%%%%%%%%%
%%                                          %%
%% Enter short notes here                   %%
%%                                          %%
%% Short notes will be after addresses      %%
%% on first page.                           %%
%%                                          %%
%%%%%%%%%%%%%%%%%%%%%%%%%%%%%%%%%%%%%%%%%%%%%%

%%\note{Sample of title note}     % note to the article
%\note[id=n1]{Equal contributor} % note, connected to author

\begin{abstract} 
The need for high-throughput, precise, and meaningful methods for measuring behavior has been amplified by our recent successes in measuring and manipulating neural circuitry.  The largest challenges associated with moving in this direction, however, are not technical but are instead conceptual: what numbers should one put on the movements an animal is performing (or not performing)?  In this review, I will describe how theoretical and data analytical ideas are interfacing with recently-developed computational and experimental methodologies to answer these questions across a variety of contexts, length scales, and time scales.  I will attempt to highlight commonalities between approaches and areas where further advances are necessary to place behavior on the same quantitative footing as other scientific fields.
 \end{abstract}

\maketitle
%%%%%%%%%%%%%%%%%%%%%%%%%%%%%%%%%%%%%%%%%%%%%%
%%                                          %%
%% The keywords begin here                  %%
%%                                          %%
%% Put each keyword in separate \kwd{}.     %%
%%                                          %%
%%%%%%%%%%%%%%%%%%%%%%%%%%%%%%%%%%%%%%%%%%%%%%

%\begin{keyword}
%\kwd{ }
%\end{keyword}

% MSC classifications codes, if any
%\begin{keyword}[class=AMS]
%\kwd[Primary ]{}
%\kwd{}
%\kwd[; secondary ]{}
%\end{keyword}

%%%%%%%%%%%%%%%%%%%%%%%%%%%%%%%%%%%%%%%%%%%%%%
%%                                          %%
%% The Main Body begins here                %%
%%                                          %%
%% Please refer to the instructions for     %%
%% authors on:                              %%
%% http://www.biomedcentral.com/info/authors%%
%% and include the section headings         %%
%% accordingly for your article type.       %%
%%                                          %%
%% See the Results and Discussion section   %%
%% for details on how to create sub-sections%%
%%                                          %%
%% use \cite{...} to cite references        %%
%%  \cite{koon} and                         %%
%%  \cite{oreg,khar,zvai,xjon,schn,pond}    %%
%%  \nocite{smith,marg,hunn,advi,koha,mouse}%%
%%                                          %%
%%%%%%%%%%%%%%%%%%%%%%%%%%%%%%%%%%%%%%%%%%%%%%

As modern techniques for recording and manipulating neural circuits have expanded our toolbox for deconstructing the molecular and cellular components of animals' nervous systems, an accompanying realization has gradually developed: to more fully comprehend the function of neural circuits and the computations underlying them, we must understand their output in an accordingly precise manner \cite{Anderson:2014ds,Krakauer:2017ie}.  Specifically, we need to measure behavior.  More careful measurements of the actions animals perform is key not just for advancing our basic understanding of nervous system function, but also in our assessment and categorization of psychiatric disorders and the development of brain-machine interfaces \cite{Wang:2014iy,Lebedev:2017bs}.  But what type of behavior do we want to measure, and once we decide on this, how do we measure it?  

Answering these questions has proven difficult, but this is largely due to conceptual limitations rather than technical ones.  If watching an animal behave, what are some precise, yet manageable, numbers we should use to describe its movements?  Is it the center of mass motion of the whole animal?  The position and velocity of the organism's body and limbs?  The dynamics of individual myosin motors within muscle tissue?  A more coarse-grained measure related to the animal's ``intended" action?  A collective variable describing the combined dynamics of many animals?  And how do we connect these scales to make inferences from the cellular and the molecular up to the movement of a limb, a wing, a finger, or an eyebrow?  This is the dilemma that those of us who attempt to measure behavior commonly face.  

While selecting the proper representation for one's measurements is hardly a problem exclusive to the study of animal behavior (e.g. ``more is different" and not wanting to ``model bulldozers with quarks" \cite{Anderson:1972ez,Goldenfeld_Simple_1999}), it is felt acutely by researchers in this field due to the multi-scale and distributed dynamics inherent to almost any behavioral process.  Cognition or sensation acts to drive muscles that drive joints that drive limbs that drive locomotion or other motions, which then send feedback signals in the reverse direction, and the cycle continues.  Where in this loop do we define behavior?  Or is it the whole loop?  And what numbers should we use to describe the observed dynamics?  These are the questions that I will focus on here, asking how to best represent behavioral data in a manner that bridges length and time scales, highlighting particularly fruitful approaches. 

Before progressing, though, it should be noted that there has been a recent proliferation of review articles discussing behavior, detailing concepts ranging from computational techniques for measuring behavior \cite{Dell:2014ho,Branson:2016ji,Robie:2017bq,Calhoun:2017bj} to finding simplicity in ``big behavioral data" \cite{2011PNAS..108S5565S,GomezMarin:2014ct} to the advent of computational psychiatry and measuring emotional states \cite{Montague:2012cp,Anderson:2014fd,Anderson:2016jv,Wang:2014iy,Huys:2016jw,Stephan:2012il} to the need for measuring behavior in the first place \cite{Anderson:2014ds,Krakauer:2017ie} to the reproducability and robustness of said behavioral measures \cite{2011PNAS..108S5537L,Fonio:2012et}.  While there will inevitably be a great deal of overlap between this review and those that have come before it, here I will focus less on the practical aspects of behavioral quantification and more on the consequences of the representational choices one makes, highlighting areas where further progress is required.

\section*{Measuring behavior on the organismal scale}

Beyond being a mere technical inconvenience, the relative lack of a quantitative language for measuring behavior has shaped the types of questions we have been able to ask.  In a laboratory setting, behavioral experiments have usually been designed to observe a restricted set of actions within the scope of a restricted environment \cite{ALTMANN:1974uj,martin2007measuring}.  To wit, the behavior measured in most of these experiments is typically performed within a ``paradigm" -- with the accompanying implication that we have tuned the animal to our quantification scheme rather than the other way around.  Examples of this approach would be placing an animal in a maze where it can only turn left or right or head-fixing a rodent, where it is asked to perform a whisking-based detection task.  While this reliance on non-naturalistic behavior sometimes emerges from a culture of treating behavior as a read-out variable of the neural hardware in question, more commonly, it is driven by an understandable desire to have a repeatable measurement that can generate high-throughput data while recording activity from neurons.  Nevertheless, the end result is to measure an over-constrained behavior that likely lies outside of an animal's typical repertoire of actions.

To move forward with the analysis of more natural behavior, we can try to imagine the best case scenario, ignoring all of the technical worries.  If we have an arbitrarily large amount of high quality data from an animal behaving with minimal artificial constraints, how should we describe it quantitatively?  To an extent, the answer here is the same as in most other scientific measurements: we desire \emph{consistency} (repeatable results), \emph{fidelity} (describing the system as accurately and completely as possible), \emph{interpretability} (ease of relating the found numbers to their biological underpinnings), and \emph{scalability} (requiring minimal manual labor or scoring without impractically taxing computational or human resources).  

While consistency and scalability can be theoretically obtained independently of the other two, fidelity and interpretability are, by definition, in tension, with measurements typically being more understandable but less accurate as we remove details.  Our goal for measuring behavior, then, is to find descriptive representations of these multi-scale processes that are as parsimonious as possible.  This trade-off naturally suggests a continuum of solutions, and in the rest of this article, we will see how researchers have represented behavioral data in varying ways, tying-together the length and time scales of naturalistic behavior at many levels of abstraction. 

\section*{Selecting a representation}

A good place to start investigating behavioral representations is to note the options available to researchers a decade ago if they wished to measure ethological behavior at the organismal scale.  One option would have been the previously mentioned paradigmatic approach, where the quantification is ingrained into the experimental apparatus itself.  Quantifying behavior in this manner has the advantages of high-throughput and consistent measurements, but it captures a very low-dimensional and potentially unnatural measurement \cite{Gao:2015gu}.  

Another approach would be to measure a coarse, yet non-paradigmatic, variable such as mean velocity or the fraction of time moving (including the laser-crossing experiments that are typical in circadian rhythm studies \cite{2002Natur.417..329P}).  These measurements are more naturalistic than paradigmatic ones, while allowing for a similar level of throughput, however, they only capture dynamics at a single scale.  This is a plausible approach when studying the effects of genetic manipulations on sleep-wake cycles, but it not be able to capture, say, the precise grooming patterns of an animal or other movements that are unlikely to be apparent by treating the animal as a point moving through space.

Alternatively, if a researcher desired a richer description of an animal's behavior, they could have developed a human-defined classification system for an animal's behavior that was then scored by a trained observer.  While providing a great deal more description, this approach is extremely labor-intensive, often requiring significant effort to devise the scoring scheme, followed by potentially months of researcher-hours to apply it.  Moreover, although the scheme one uses can be elaborated in detail, there will inevitably be user-specific variability in its application.  More problematic, since behaviors are defined and delineated intuitively, it is difficult to quantitatively argue that one individual's or group's representation of the behavior is more accurate or appropriate than another's, further limiting reproducibility.  Lastly, this approach implicitly assumes that behavior can be described in terms of hopping between discrete states without showing, from the data, that such a model is indeed a reasonable representation in the first place.

Skipping ahead to the present, all three of these options are still frequently used, often generating novel insights into behavior and the mechanisms driving it.  Automation has greatly increased the throughput of the first two described options, especially in small organisms like worms \cite{Yemini:2013bd,churginlongitudinal}, flies \cite{2015PNAS..112.6706A,Branson:2009jf}, and zebrafish larvae \cite{PerezEscudero:2014ed,Naumann:2016ck,Orger:2017jw}.  Moreover, supervised machine learning techniques have greatly improved the repeatability and decreased the manual effort required to analyze behavioral data with user-defined classification of behavior \cite{Dankert:2009fa,Kabra:2013jk,deChaumont:2012du,2013NatCo...4E1910K,2015PNAS..112E5351H}.  That being said, the fundamental difficulties with these approaches remain - the level of behavioral description is either coarse-grained or behaviors are intuitively defined, explicitly encoding a human observer's underlying assumptions about an animal's behavior.  Thus, we come to a fundamental query: how can we leverage modern data-collection techniques to extract complex behavioral representations in manner that is transparent and repeatable, with explicitly-stated and testable assumptions that shed light onto particular biological questions?  The answer to this requires thinking about the general principles.

 \begin{figure}
	\includegraphics[width=.9\columnwidth]{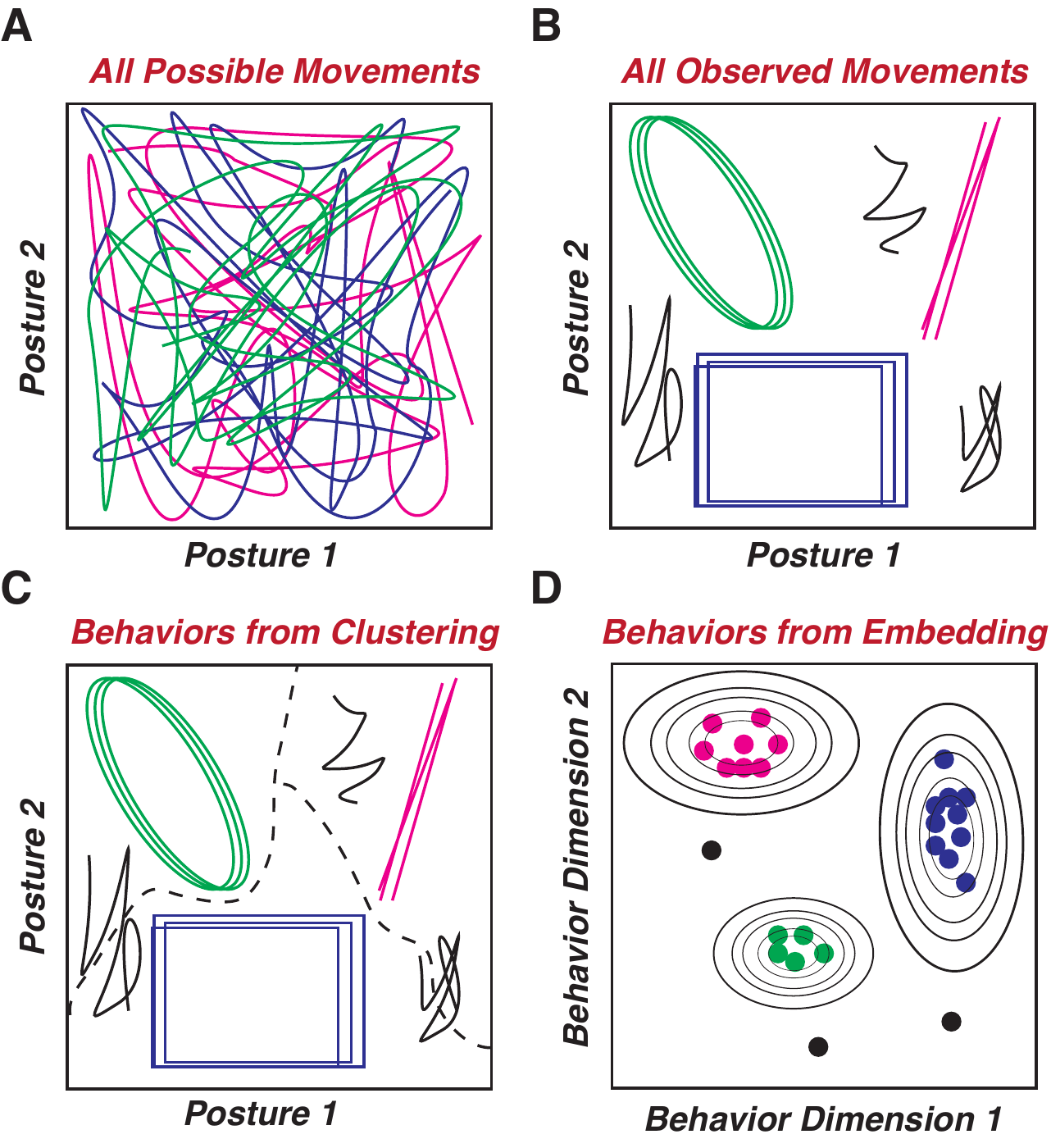}
	\caption{\textbf{Approaches for identifying stereotyped movements.}
		\textbf{A.} Representation of all of the movements an animal could theoretically make.  For instance, each line could be the dynamics of two joint angles, say, the bending of a knee and an ankle, or another set of postural variables over time.  Although an animal could potentially move with any of these postural trajectories, many of the motions here would be only rarely performed.  \textbf{B.} How we observe most animals to move.  Specifically, they use a relatively small portion of their potential behavioral repertoire (stereotyped behaviors, colored lines) along with a few instances of less-robustly observed ones (non-stereotyped behaviors, black lines).  \textbf{C.}  One way to isolate stereotyped behaviors is to break-up the observed trajectories into clusters (denoted by dashed lines).  \textbf{D.} An alternate means of identifying stereotyped behaviors is to transform the dynamics in such a way that, for instance, each time one of the trajectories in \textbf{B} is performed, a dot is placed using a low-dimensional embedding to a different space.  Similar trajectories are mapped near each other (dots), and stereotyped behaviors could be identified as peaks in the density contours (lines) of this map.}\label{fig1}
\end{figure}

\section*{Stereotypy as a general principle}

One area of organismal-scale research where the challenges in measuring behavior have become predominantly technical is biolocomotion, the study of how animals move through their environments \cite{2000Sci...288..100D,childress2012natural,2006SIAMR..48..207H,Miller:2012fv}.  Here, while many deep questions regarding the performance, control, and evolution of these behaviors remain to be answered, there is a generally-agreed-upon framework for measuring behavior: most researchers study dynamic trajectories of motion, typically center-of-mass, body bending, and/or limb trajectories.  What is it about biolocomotion that has made it amenable to this type of agreed-upon representation?

Part of the reason for this advantage is the clear ethological context of the actions studied -- moving from one place to another quickly, efficiently, and robustly.  Thus, there is a natural mathematical formalism to translate between scales, namely Newtonian mechanics, and the behaviors in question are clearly separable from other actions that the animal performs.  Even in cases where the mathematics underlying this translation are untenably difficult to analyze directly, robots can serve as the physical equivalent of generative models to bridge this gap \cite{McInroe:2016wk,aguilar2016review,Dickinson:1999ty}.  Moreover, concepts such as optimal control or energy-efficiency provide a theoretical basis for providing meaning to the investigations \cite{Cowan:2014jv,Roth:2014cl,alexander1996optima}.  Another factor is that these behaviors are highly stereotyped, with physical constraints typically allowing for only a small number of movement patterns or gaits \cite{Full:1999vc}.  Although animals are capable of moving their limbs in an extremely large manner of ways \cite{mpfc}, during locomotion, their motion lies on a very low-dimensional set of postures, with small perturbations either corrected for or used to actuate control \cite{Ristroph:2010jk,Jindrich:2002tr,Revzen:2012cj}.  

Inspired by these studies, much of the recent progress in developing tools for data-driven and unsupervised (i.e. without the aid of human-labeled examples) analysis of animal behavior has resulted from this observation that a large fraction of animal movements are low-dimensional compared to the animal's total capacity for movement and are often repeated in a similar manner (Fig. \ref{fig1}) \cite{Osborne:2005bj,2011PNAS..108.7286S,2011PNAS..108S5565S,Berman:2014ef}.  However, in order to proceed, we must have a more precise mathematical description of stereotypy (i.e. defining what we mean by ``low-dimensional," ``similar," and ``movement").  The goal here is to put the human at the beginning of the analysis process (defining stereotypy) as well as at the end (interpreting behavioral outputs of the analysis process), rather than in the middle, as is the case for label-based, or supervised, methods for behavioral analysis.

Several recent studies have developed tools for finding these stereotyped behaviors across a range of model organisms during (relatively) free behavior, from worms to rodents \cite{Stephens:2008dk,Brown:2013ew,Berman:2014ef,Wiltschko:2015ho,Kato:2015ck,2017PhBio..14a5002T,Klibaite:2017el}.  Although these researchers have all taken widely-differing technical approaches, there are key similarities that join their efforts together.  The common logic between these methods points toward a shared definition of stereotyped behavior and forces us to ask a pair of fundamental questions: what does it mean for two behaviors to be similar or different, and how do we place a number on this difference?  One thing that is important to note, though, is that none of the approaches described below are strictly \emph{unbiased}, despite the term being often brandished when describing their advantages.  The implication in calling these unsupervised approaches unbiased is typically that the analyzer is removing themselves fully from the loop.  Each choice a researcher makes, though, has consequences, regardless of how explicit those choices are, but the key to all of these approaches is that the consequences of these options are readily apparent.  

\section*{Finding stereotyped movements in behavioral data}

\begin{figure}
  \includegraphics[width=.9\columnwidth]{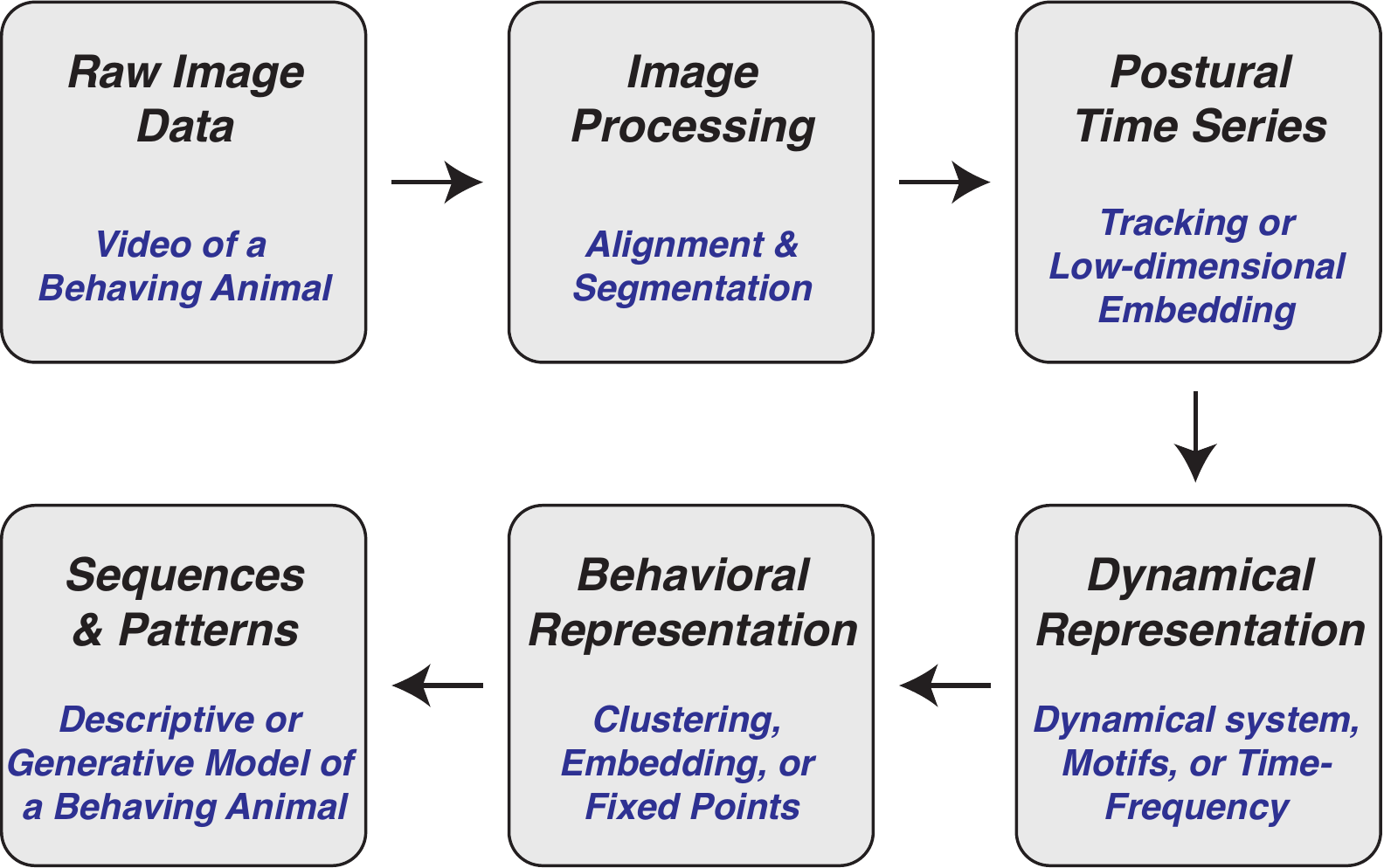}
  \caption{\textbf{Archetypical data analysis pipeline for identifying stereotyped behaviors automatically from data.}}\label{fig2}
  \end{figure}

Although superficially distinct, there are surprising similarities in the underlying bases of different approaches for automatically identifying stereotyped behavior from videos (we will ignore other modalities for the moment) of freely-behaving animals.  The general framework has been to first extract a low-dimensional \emph{postural time series} from a data set, followed by a translation of these postures into a \emph{dynamical representation} that is used to create a \emph{behavioral representation} that isolates individual stereotyped actions.  If desired, an animal's dynamics within this behavioral representation can be observed over time, finding patterns and sequences of behavior (Figure \ref{fig2}).

\subsection*{\textbf{Extracting postural time series}}
The first step in almost any of these analyses is to isolate the animal's posture from the raw video data.  Here, by posture, I mean a measure that describes the configuration of an animals' body and limbs at a given point in time (describing how they move will come in the next section).  Usually, we prefer to describe this configuration in a manner that is in the body frame of the animal so that behavior is measured independently from spatial position or orientation.  It is from this snapshot that further analyses will be devised, and it is here where organism-specific practicalities are most apparent.  

\begin{figure*}
  \includegraphics[width=2\columnwidth]{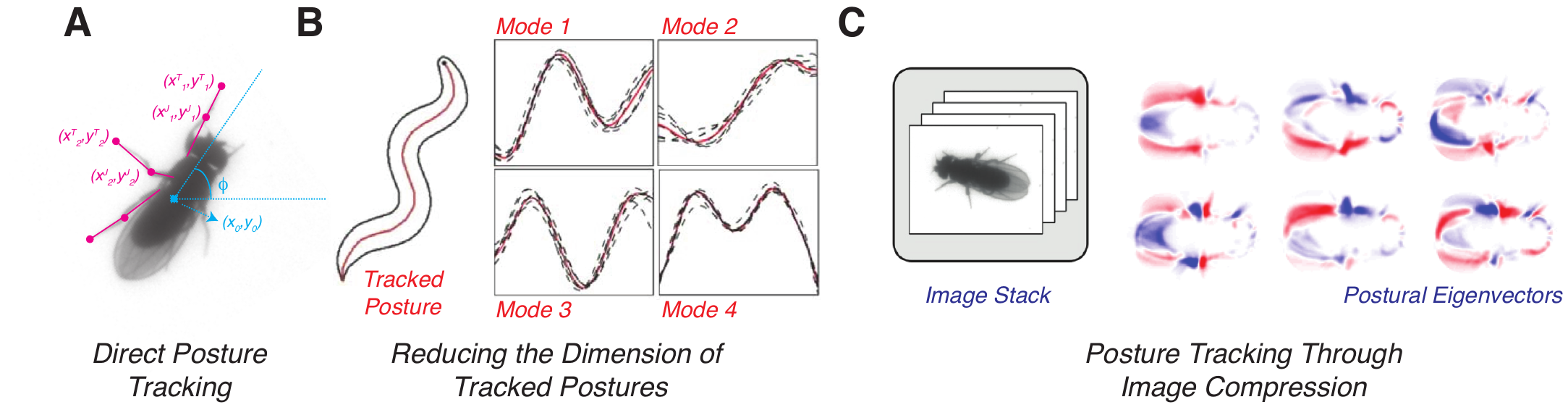}
  \caption{\textbf{Examples of postural representations.}  \textbf{A.}  A schematic for how posture is typically represented by assigning body frame coordinates, here for a fruit fly.  This assignment is usually created from manual tracking or machine vision techniques.  \textbf{B.} Using variations in the tracked centerline of the nematode \emph{C. elegans} (left) to find a set of postural modes (right).  Here, principal components analysis is used to find a set of ``eigenworms," where the original centerline can be largely reconstructed through a linear combination of these centerline variations (adapted from \cite{Stephens:2008dk}).   \textbf{C.}  In cases where tracking is not feasible due to occlusions, high-dimensionality, and/or large data sets, an alternative approach has been to use image compression to find postural modes, such as those seen in the fly images here (adapted from \cite{Berman:2014ef}).  Here, red and blue represent positive and negative eigenvector magnitudes, respectively, that are the result of concentrating as much of the data's variance in as few directions as possible.  The original image can be reconstructed via a linear combination of all the modes plus an overall mean, and time series can be generated by observing sequential images' projections onto these postural modes.}\label{postural}
  \end{figure*}

This latter point can be readily seen in the difference between describing a nematode like \emph{C. elegans} and a fly like \emph{D. melanogaster} (Figure \ref{postural}).  While almost all of the dynamics of worm behavior could be described by the motion of its centerline, a fly's movement is the combination of six legs (each with two joints), two wings (each capable of moving with three degrees of freedom), and other body movements such as abdomen bending.  These body plans clearly require different representations, even if we expect both to be relatively low-dimensional over the course of typical activities the animals perform.  A rodent, with a more flexible body, or a human, with its typical bipedal walking gait, would require different representations still.  In all cases, though, the aim is to take a high-dimensional measurement -- say, thousands to millions of pixel values -- and reduce it to a low-dimensional set of numbers describing the animal's posture.

The traditional, and in some senses optimal due to its interpretability, manner to achieve a set of low-dimensional time series has been to track the positions of individual body parts such as joints, leg tips, the tail, or the head.  Outside of animals with relatively simple morphologies like \emph{C. elegans}, this is an extremely difficult computer vision problem that has been the subject of comprehensive discussions elsewhere \cite{Dell:2014ho,Robie:2017bq,Branson:2016ji}.  Even in the case of worms, new image analysis methods have been necessary to account for events where the worm crosses itself \cite{Nagy:2015ha,Broekmans:2016eh,2016PNAS..113E1074N}.  For legged animals, most automated methods typically require either attaching markers to the animal or large amounts of manual correction.  Recent advances in experimental design \cite{2013NatCo...4E1910K,Mendes:2013cu,Machado:2015jh} and computational algorithms \cite{Dell:2014ho,Ristroph:2009eb,Fontaine:2009em,Uhlmann:2017ih} provide hope for improving the state-of-the-art moving forward, but for large data sets containing up to billions of images, tracking individual body parts is not currently practical, especially for 2-dimensional images.

Instead of directly tracking, a common approach has been to think about postural decomposition as an image compression problem.  After doing some image processing to isolate the animal from the background and align it translationally and rotationally to a template image, the tactic taken by work in flies \cite{Berman:2014ef} and mice \cite{Wiltschko:2015ho} has been to perform a dimensionality reduction operation like Principal Components Analysis (PCA) on the raw image pixel data.  This process allows for images of an animal with complex morphology to be repeatably and continuously mapped into a relatively small set of time series, much like direct tracking of joint angles would do, but with vastly fewer errors and no need for manual inspection (Fig. \ref{postural}C).  This process has the disadvantage, however, of creating relatively uninterpretable time series, a fact we need take into account when moving toward a dynamical representation.

\subsection*{\textbf{Building representations of dynamical behavior}}
When defining stereotyped behavior, we typically think of movements, not postures.  For example, we wouldn't describe walking as bending the right knee at $73.1^o$, the right ankle at $15.23^o$, and so on, but rather as a trajectory of these angles through time.  As a result, to measure stereotyped behaviors, we need to create a dynamical representation that describes how the measured postural time series are changing.  Building such a representation can be achieved by either directly fitting a differential equation to the postural data or through attributing features that incorporate dynamics such as temporal motifs or time-frequency features to individual segments of the data.  We will see examples of each of these approaches momentarily.  From here, one would like to create a behavioral representation, which can be thought of as longer-time scale changes in the underlying postural movements that generate the observed postural motions. For instance, giving the relative velocities of each of an insect's six legs might be a dynamical representation, but saying that the animal is walking with an alternating tripod gait would be a behavioral representation.  Of course, we need to make this idea more precise, and we will see how several different studies have done this, each with associated strengths and challenges.

The most straight-forward process for building a dynamic representation is to eliminate the step of finding postural time series and instead create a manually-curated set of dynamical features that are later used as the input to either a supervised classifier or a clustering/embedding algorithm \cite{2014Sci...344..386V,Geng:2003tg,Ghosh:2012ch,Kabra:2013jk,2010NatCo...1E..68J,Golani:1999vt}.  While relatively easy to implement, this approach risks missing elements of behavioral dynamics not captured in the list, and each of the measurements potentially has different units (e.g. velocity, angular velocity, acceleration, distance from another animal), requiring additional conversion factors or assumptions about equal variance that could affect any analysis' outcome in subtle ways.

\begin{figure*}
  \includegraphics[width=2\columnwidth]{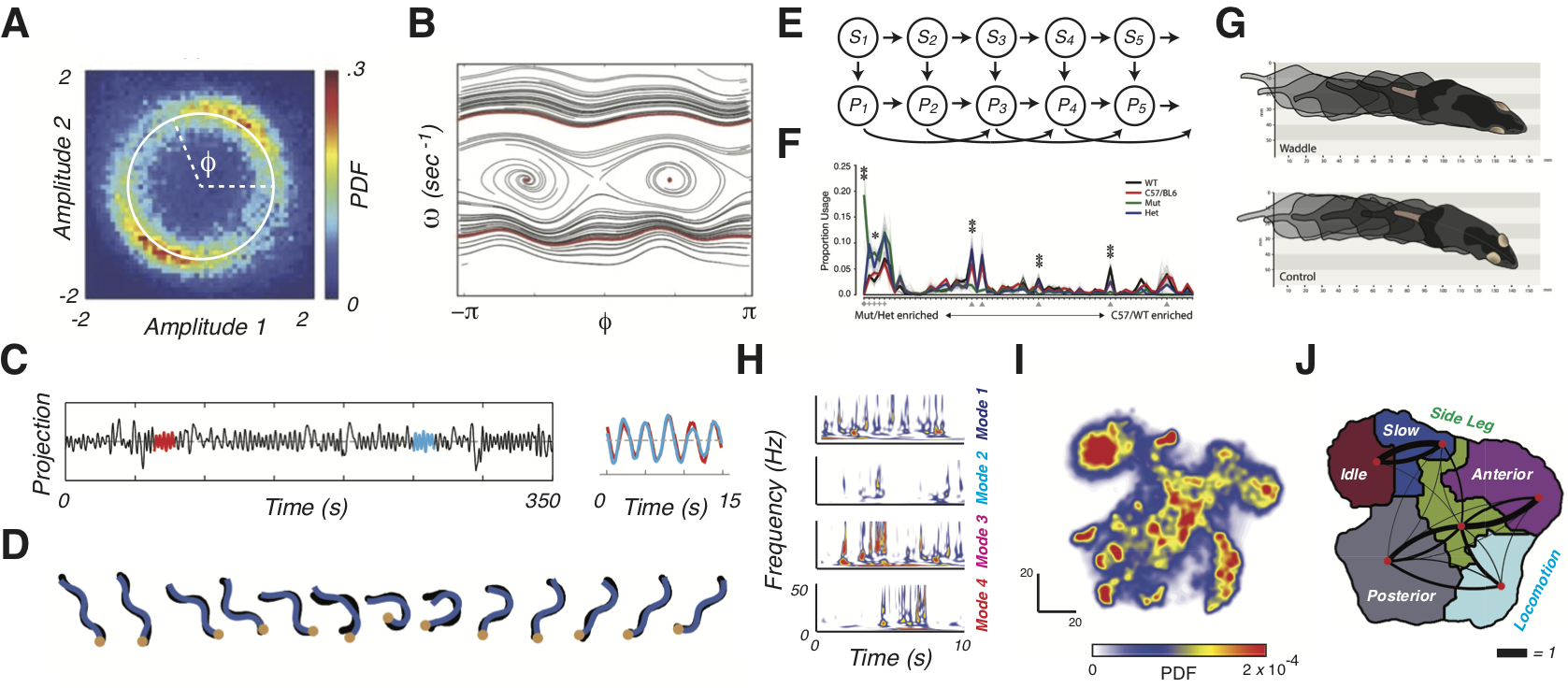}
  \caption{\textbf{Examples of dynamical and behavioral representations.}  \textbf{A.} For \emph{C. elegans}, a histogram of projections onto the first two ``eigenworms" (the left two curves in Fig. \ref{postural}B) shows a low-dimensional structure that can be parameterized by a single phase variable, $\phi$.  \textbf{B.} Fitting the dynamics of this variable to a deterministic dynamical system yields this phase map, with forward and backward locomotion naturally emerging as traveling wave trajectories at the top and bottom, respectively, and two fixed points in the middle corresponding to two different pause states (\textbf{A} and \textbf{B} are adapted from \cite{Stephens:2008dk}).  \textbf{C.} An alternative approach to represent \emph{C. elegans} behavior is via motif-finding.  Here, time-series of projections onto the eigenworms are scoured for repeated patterns (e.g. the blue and red curves here).  These patterns are then catalogued and used as the basis for a behavioral representation (adapted from \cite{Brown:2013ew}).  \textbf{D.}  Instead of using dynamical motifs directly, the worm's behavior can be captured as a sequence of postures, as seen in this example from \cite{GomezMarin:2016db}.  \textbf{E.}  The approach taken by \cite{Wiltschko:2015ho} was to fit an autoregressive hidden markov model (AR-HMM) to postural data of mouse movements, generated in a similar, but not identical, manner to that seen in Fig. \ref{postural}C.  Here, each $P_t$ is a vector of the animal's postural mode values at time $t$, and $S_t$ is an underlying state that affects the dynamics of postural outputs.  Here, arrows imply direct dependence (i.e. $P_t$ is a stochastic function of $S_t$, $P_{t-1}$, and $P_{t-2}$, and so on).  It is assumed that the time scale for changes in $P$ is much faster than that for changes in $S$.  This latter time scale, a parameter in the model, sets the distribution for the length of time that an animal stays within a particular behavioral state.  \textbf{F.}  Average behavioral usage frequencies using an AR-HMM for four different mouse genotypes: Wild type, C57/BL6, as well as homozygous (\emph{Mut}) and heterozygous (\emph{Het}) mutations in the aretinoid-related orphan receptor 1$\beta$ (Ror1$\beta$) gene.  \textbf{G.}  Distinct walking gaits found in the \emph{Mut} (top) and C57/BL6 (bottom) mice (\textbf{E-G} adapted from \cite{Wiltschko:2015ho}).  \textbf{H.} An example of a time-frequency analysis representation from freely-moving fruitflies, where each set of axes represents a mode, and the colormap values indicate the continuous wavelet transform amplitudes for at each point in time.  This approach allows for multiple time scales to enter the dynamical representation.  \textbf{I.} Probability density resulting from embedding points into 2-d such that two instances when a fruit fly is moving similar parts of its body at similar speeds are mapped nearby.  Note the peaks and valleys.  Here, the peaks represent stereotyped behaviors. \textbf{J.} Break-down of the behavioral representation in \textbf{I}, with names for the behaviors within each of these regions manually labelled.  Black lines are proportional to the transition probability between moving from one coarse region to another, with right-handedness implying the direction of transmission.  (\textbf{H-J} adapted from \cite{Berman:2014ef,Berman:2016db}).}\label{behavioral_representation}
  \end{figure*}

Ideally, an appropriate dynamical representation would emerge naturally from postural dynamics.  To date, the clearest example of using postural data to explicitly generate a dynamical system that provides a natural behavioral representation is the work on \emph{C. elegans} from Stephens et al \cite{Stephens:2008dk,2011PNAS..108.7286S}.  Here, the authors found that the majority of a worm's motion can be described by the progression of a single phase variable that can be thought-of as the advance of a traveling wave moving up or down the animal's body.  Fitting the observed dynamics of this variable to a model with a deterministic and a stochastic component (Fig. \ref{behavioral_representation}A-B), they find that the worm's behavior can be described as a set of dynamic attractors with switching times that are predictable from the statistics of the underlying noise.  Although applying such methods directly to higher-dimensional data sets like those generated from legged animals can be challenging, recent advances in finding dynamical models that best describe a continuous time series provide future avenues for exploration \cite{Bongard:2007ky,2017arXiv170506370N,2015NatCo...6E8133D}. 

Another approach to building a dynamical system representation is to fit a statistical model to the data.  A prominent example of this can be seen in the work of Wiltschko et al \cite{Wiltschko:2015ho}, who took collected postural time series data of mice and fit an Autoregressive Hidden Markov Model (AR-HMM) to their data (Fig. \ref{behavioral_representation}E-G).  One can think of this approach as fitting small segments (less than 1~s) to linear dynamical systems, and that the animal is switching between these systems with time scales that are significantly longer than those of the dynamics within a given system.  This method creates a dynamical representation (bottom row of Fig. \ref{behavioral_representation}E) at the same time as it creates a behavioral representation (top row of Fig. \ref{behavioral_representation}E).  

While the ability to simultaneously represent dynamics and behavior is a distinctive advantage of the AR-HMM approach, it is also a limitation since it requires a parameter that sets the overall time scale of staying in a particular behavioral state.  One could imagine amending this limitation by adding additional time scale parameters when fitting the model, but this still requires a hand-tuning of the time scales available to the system, as well as a corollary assumption that the amount of time an animal spends in a particular behavior must follow an exponential distribution.  The time spent performing a behavior, however, can range over orders of magnitude  -- from a reflex lasting tens of milliseconds to a night's sleep -- and long time-scale dynamics are often observed in behavioral data \cite{2011PNAS..108.7286S,Berman:2016db}. Moreover, if one wishes to directly measure the time scales evident in a particular data set, the fact that the approach used relies on this type of structure as an assumption can be confounding.

A complementary approach is to create a multi-scale dynamical representation that forms the basis for a behavioral representation.  This can be achieved through finding motifs of varied lengths in a data set \cite{Brown:2013ew,Schwarz:2015bk,GomezMarin:2016db} or using a time-frequency analysis approach like a wavelet transform \cite{Berman:2014ef,Klibaite:2017el,Sakamoto:2009cf} to represent postural dynamics across a variety of time scales.  For the case of motif-finding (Fig. \ref{behavioral_representation}C-D), one finds postural patterns that commonly occur throughout a data set and looks for when the animal exhibits similar dynamics.  The relative frequency and patterns of use for these motifs can be used to create ``behavioral fingerprints" for individuals or collections of animals differing in genotypes, neural manipulations, or other conditions of interest.  The resulting behavioral representation is thus the set, frequency, and ordering of motifs that an animal performs.  A difficulty of this approach, however, is that results may not always be robust to slight changes in postural dynamics such as changes in frequency or relative phasing between limbs. 

An alternative approach to capture behavioral dynamics across multiple time scales is to use time-frequency analysis (Fig. \ref{behavioral_representation}H).  Here, one takes the set of postural time series, determines a wide range of frequencies that are present in each time series and measures the relative importance of each of these frequencies as a function of time.  This importance is often quantified via a wavelet transform \cite{debnath2002wavelet}, which uses a trade-off between accurate temporal resolution and poor frequency resolution at high frequencies and poor temporal resolution and accurate frequency resolution at low frequencies to generate a multi-scale representation of the animal's postural movements.  The resulting dynamical representation for a single point in time is thus a set of wavelet amplitudes for a collection of frequencies from each of the observed postural time series.  Despite the fact that the wavelet transform contains both amplitude and phase information, it is typical to only use the amplitude information, as this eliminates many of the robustness issues experienced in the motif-finding case.  Behavioral representations can then be obtained from either clustering \cite{2017PhBio..14a5002T,Sakamoto:2009cf} or low-dimensional embedding (Fig. \ref{behavioral_representation}I-J) \cite{Berman:2014ef,Klibaite:2017el} of the resulting vector of amplitudes.  Typically,  when embedding these feature vectors, an anisotropic density across this space emerges, with local peaks corresponding to particular stereotyped behaviors.  Accordingly, one could treat the behavioral representation as either the density itself or the sequence of peaks that an animal visits.

\subsection*{\textbf{Discrete vs. continuous behavioral representations}}

Note how we now have seen that behavioral representations can either be discrete (e.g. clusters or motifs) or continuous (e.g. densities or non-piecewise dynamical models) and that discrete representations can often be derived from continuous ones (e.g. fixed points or peaks).  So which is better?  Ideally, one is able to identify a discrete representation through the fixed points of a dynamical model, but this is not currently practicable for animal morphologies more complicated than a worm's.  For other systems, though, like most methodological questions, the answer depends on the experimental exigencies at play, and performing both often provides additional context and information.  

On one hand, in favor of continuous representations, it is more intellectually satisfying to show that a discrete representation arises naturally out of a data set without imposing such a structure a priori.  Even in the case where a discrete representation is appropriate, it may be that the interesting measurements to note are the subtle distinctions on the edge of the peaks.  Additionally, although many of the movements an animal performs are stereotyped, not all of them need to be.  An important aspect of continuous representations is that they allow for the ability to have portions of time where the animal is performing non-stereotyped dynamics (i.e. they do not remain stationary on the map in Figure \ref{behavioral_representation}J).  Results from fruit flies show that the animals perform non-stereotyped behaviors approximately half of the time \cite{Berman:2014ef,Klibaite:2017el}, implying that one must be careful when interpreting a representation that places all time points into a cluster.

On the other hand, though, if the data indeed has clusters, one should perform clustering in the high-dimensional space that retains all of the information in the data and where partitioning algorithms are more likely to succeed  and one does not have to worry about the specific form of the length-scale distortions that any nonlinear embedding necessarily creates \cite{2017PhBio..14a5002T}.  However, while formalisms such as AR-HMM allow for the building of a type of dynamical model, they also rely on underlying assumptions about a single time scale that could over- or under- partition the data.  Accordingly, researchers need to think carefully about the consequences of these choices of representation and tailor their approach to the questions at hand.

\section*{Future Challenges}
Many of the next steps in building representations for measuring behavior involve building representations that link postural dynamics to dynamics of other variables, including space, other behavioral modalities, other individuals, and neural dynamics. 

\subsection*{\textbf{Joint representation of space and posture}}
An interesting observation about almost all of the representations in the previous section is that the typical quantities measured in coarse behavioral assays, namely spatial position, orientation, and their derivatives, are the first aspects to be eliminated.  This is performed to ensure that one measures motions in an animal's own frame, but there are numerous scenarios in neuroscience and social behavior where we would like to look at the interactions between location, movement, and behavioral patterns, ideally generating a joint representation.  

A natural question here is, why not simply add the postural dynamics as an extra time series to be thrown-into one's favorite behavioral mapper or classifier?  The difficulty here is that the variables describing dynamical representation -- derivatives or spectral transforms of joint angles or postural modes -- all have the same units, and these units differ from those of the spatial variables.  Thus, a unit conversion must occur, requiring at least one arbitrarily-chosen parameter.  

Current solutions have been to measure behavior conditioned on position or position conditioned on behavior \cite{Klibaite:2017el,Wiltschko:2015ho,benjamini2011quantifying} or to measure a response field averaged across individuals \cite{Katz:2011fb}, but this  does not provide a true joint representation.  As an example, if one animal performs the exact same motion twice, but in slightly different locations, are those two behaviors closer or further away than the animal performing two slightly different motions but at the exact same position?  Finding systematic and precise quantifications to answer this question (and the answer might change depending on the precise scientific investigation at hand) will be key to building joint positional-postural representations.

\subsection*{\textbf{Collective and social behavior}}
Similar to the difficulty of representing space and posture simultaneously, we face a problem when attempting to describe the collective dynamics of many individuals moving together.  This is often achieved through measuring an order parameter that is related to the proportion of individual velocities pointed in the same direction \cite{1995PhRvL..75.1226V,Toner:1998vc,Buhl:2006fp}.  Ideally, though, one would like to capture metrics that describe the collective dynamics of many individuals in a manner as rich as the previously-described approaches for single animals.  Particularly fruitful ideas here borrow techniques from fluid dynamics, including the use of Lagrangian coherent structures \cite{doi:10.1146/annurev-fluid-010313-141322} and dynamic mode decomposition \cite{Brunton:2016ga} to generate continuum-based models of many organisms moving collectively.

An additional challenge arising in social behavior is that much of the research described previously focuses on the physical motion of an animal's limbs and body, but in the case of social interactions, capturing other aspects of behavior such as the production of audio and substrate-borne signals will be necessary to fully describe the animals' dynamics.  There have been many recent successes relating behavioral dynamics to, for example, audio dynamics through asking what behavioral features predict the performance of a particular song or song type using methods such as general linear models (GLMs) \cite{2014Natur.507..233C,LaRue:2015cm,Neunuebel:2015bm}, and improvements in automated methods have increased the throughput of audio data analysis \cite{Arthur:2013di,Tabler:2017hn,Mets:2017gu,VanSegbroeck:2017ic}.  Ideally, though, we would be able to create a joint representation of the alternative behavioral modalities and the postural movements occurring at the same time that more fully links the dynamics of these processes.  

\subsection*{\textbf{Linking neurons to behavior}}
As our ability to record neurons in freely-behaving animals increases, the need to represent neural activity jointly with behavior is becoming increasingly apparent.  As with multi-modal dynamics, most current approaches to neuro-behavioral analysis \cite{Robie:2017bz,2014Sci...344..386V,Clemens:2015jl,Wang:2016cr,Billings:2017dq,Gepner:2015fp,Kato:2015ck,Dunn:2016gr} take a correlative or decoding approach: given one knows something about neural dynamics, what can one predict about behavior, or vice versa?  This could take the form of ``given a neural stimulation what did the animal do?" or ``is it possible to predict an animal's behavior from neural dynamics?"  While these are necessary first steps toward building our understanding of how neural circuits drive behavior, to more fully comprehend the interplay between these circuits and how behavior feeds back onto neural responses, we need to devise methods to simultaneously analyze the combined dynamics of posture and neural activity.  

One potential avenue for achieving this aim is to combine experimentally-tested computational models of neural dynamics with high-resolution behavioral measurements and perturbations.  Ideas toward this end have been put forward in the nascent field of computational psychiatry, where neural models, ranging in scale from small collections of neurons to individual brain nuclei to whole brain dynamics \cite{Wang:2014iy,brodersen2014dissecting,Huys:2016jw,Muldoon:2016bg} are manipulated or systematically controlled to see how system-wide outputs are affected.  Although in these studies, outputs are usually measured in terms of neural activity alone, a joint representation of behavioral outputs in model organisms or human patients and model-specific control dynamics of the neural circuit present an intriguing path forward.  This would also allow ideas from control theory to inform the discussion \cite{Cowan:2014jv}, building a framework where feedbacks between neural activity and behavior could be more thoroughly linked.

\section*{Toward theories of behavior}
The previous point about the use of theory in generating behavioral representations brings us back to the beginning of our discussion.  Our fundamental challenges still remain at the conceptual rather than the technical level.  Despite the significant advances in measuring behavior over the past few years, the ultimate goal of these approaches -- understanding how and why animals control and generate particular sequences of physical movements -- requires developing theories and models to serve as connective tissue, providing context and justification for the measurements we make and allowing us to make predictions that suggest future experiments.

But what should these theories look like?  Does this mean we should turn behavior into particle physics?  What is the atom or proton or quark of behavior?  Does it even make sense to discuss behavior as if there is a set of underlying first principles from which all actions are derived?  Like most questions in biology, we can begin to make progress by looking to evolution.  Specifically, we cannot forget that almost every behavior has a goal: to increase an animal's probability of passing its genes to subsequent generations.  Thus, all movements are placed in the context of how they aid in the performance of one or many tasks.  

This viewpoint, shared by many of us who refer to ourselves as ``computational ethologists" \cite{Anderson:2014ds} (whether this is different than ``ethologists with fancy computers" is a discussion for a different article), makes an argument to engage in a parallel endeavor to the mapping and manipulation of neural circuits.  We should search for what Richard Dawkins referred to as ``software explanations of behavior" \cite{Dawkins:1976ti}.  The most famous example of this type of analysis is Tinbergen's hypothesis that animals' behavioral drives can be explained via a hierarchically-organized set of competing impulses, based on both observations and ideas about optimality and evolvability \cite{Tinbergen:1951tm}.  This idea, independently developed by Herbert Simon in the context of engineered systems \cite{Simon:1962vg,simon73}, provides testable consequences that have lead to further investigations and theories across a wide variety of systems \cite{Dawkins:1976wx,lefebvregrooming,Berman:2016db,Solway:2014ij}.  Similarly, ideas about optimizing feedback control and energy-efficiency have shaped biolocomotion studies \cite{sponberg2017}, and concepts from reinforcement learning have served as a starting point toward investigations into the neural implementation of learning \cite{Montague:2012cp,Niv:2009dw}.  

In each of these examples, observations about behavior have been used to make inferences about the brain's functioning that do not explicitly rely on detailed models or knowledge of brain dynamics or morphology, potentially providing general principles that apply across systems.  When deciding what type of behavior to measure and how to measure it, we either intentionally or unintentionally rely on theories such as these when we choose a behavioral context, select length and time scales, or decide how to analyze the data.  

Only through consciously generating and interacting with broad theoretical concepts can we create a fuller understanding of how neural systems function to produce movement and behavior.   For example, the idea of using stereotyped movements as a scale for behavioral measurements builds upon observations about low-dimensionality in movements and the commonality of neural circuitry such as central pattern generators devoted to the performance of periodic activities.  Taking these assumptions directly into account has allowed the methods discussed in this review to be developed, and the identification of further concepts will be essential to their expansion, refinement, and application.  At its core, ``What type of behavior do we want to measure?" is a question that relies on theoretical insight for its answer, and future efforts toward quantitatively linking behavior to its physiological underpinnings will greatly benefit from approaching experimental design and analysis accordingly.

\section*{Competing interests}
  The author declares no competing interests.

%\section*{Author's contributions}
%    Text for this section \ldots

\section*{Acknowledgements}
The author would like to thank Avani Gadani, Alex Gomez-Marin, Itai Pinkoviezki, Jennifer Rieser, Carlos Rodriguez, and Kun Tian for comments and suggestions on the manuscript.  This work was partially supported by NIMH 1R01MH115831-01.

%  Text for this section \ldots
%%  After submission of the .TEX file,                     %%
%%  you will be prompted to submit your .BBL file.         %%

% if your bibliography is in bibtex format, use those commands:
\bibliographystyle{vancouver} % Style BST file (bmc-mathphys, vancouver, spbasic).
\bibliography{review.bib}      % Bibliography file (usually '*.bib' )

\end{document}